# Greenhouse Gas Emissions and its Main Drivers: a Panel Assessment for EU-27 Member States

Ionuț Jianu[1], Sandina-Maria Jeloaica[2], Maria-Daniela Tudorache[1]

[1](Department of Economics and Economic Policies, Bucharest University of Economic Studies, Romania)
[2](Faculty of Finance, Insurance, Banking and Stock Exchange, Bucharest University of Economic Studies, Romania)
*Corresponding Author: Ionuț Jianu[1]*

**ABSTRACT:** This paper assesses the effects of greenhouse gas emissions drivers in EU-27 over the period 2010-2019, using a Panel EGLS model with period fixed effects. In particular, we focused our research on studying the effects of GDP, renewable energy, households energy consumption and waste on the greenhouse gas emissions. In this regard, we found a positive relationship between three independent variables (real GDP per capita, households final consumption per capita and waste generation per capita) and greenhouse gas emissions per capita, while the effect of the share of renewable energy in gross final energy consumption on the dependent variable proved to be negative, but quite low. In addition, we demonstrate that the main challenge that affects greenhouse gas emissions is related to the structure of households energy consumption, which is generally composed by environmentally harmful fuels. This suggests the need to make greater efforts to support the shift to a green economy based on a higher energy efficiency.

## I. INTRODUCTION

Climate change is the consequence of the harmful human activity that has gradually affected the environment as a result of the fossil fuels used for energy production. Gases resulting from energy production have amplified the greenhouse effect, leading to an increasing global warming. Climate change is a prominent subject nowadays and an impressive amount of studies were conducted on this issue, but there are still important unknowns and details to be established in order to better steer policy debates and to feed decision makers.

European Union is a global leader on addressing climate change mitigation and adaptation and a large number of initiatives were or are to be implemented. However, continuously monitoring and assessing their impact on the final goal (namely limiting the greenhouse gas emissions) is needed to ensure that the right track is observed. EU developed a very complex and ambitious plan for 2050 to reduce greenhouse gas emissions, namely EU Green Deal. Together with the EU, other countries, considered responsible for global pollution, set out the objective to reduce their greenhouse gas emissions, in particular the United States (until 2050), China (until 2060) and India (until 2070). Currently, China is considered the largest global polluting actor, being responsible for 28% of total global greenhouse gas emissions.

The motivation for choosing this theme consists in the importance of this subject and the climate change related risks that are evolving on an increasing path. Doing nothing may be fatal on long-term for further generation, but tackling climate change actions should be also balanced and should be implemented in an inclusive manner. Having in mind the importance of this subject and the scientific literature findings (presented in next section), the main objective of this paper is to examine the greenhouse gas emissions drivers and their effects in EU Member States over the period 2010-2019. In order to achieve the main goal of the paper, we have set the following specific objectives: (i) checking the relationship between greenhouse gas emissions and GDP per capita; (ii) examining the relationship between renewable energy and greenhouse gas emissions; (iii) checking the link between final households energy consumption and greenhouse gas emissions; (iv) examining the link between waste generation and greenhouse gas emissions.

## II. LITERATURE AND CONTEXT

Climate change and the relationship between economic growth and GHG emissions has been long debated. The extensive literature on this subject provides different assessments. On the one hand there is the optimistic approach that suggests it is possible to achieve both economic growth and GHG reduction sufficiently to limit the temperature increase to 1.5˚C (as committed by governments in the Paris Agreement), based on technological progress and the implementation of the right policies [[1], [2], [3]]. On the other hand, there is a less optimistic approach supported by theorist that consider it is not possible to reduce the GHG emissions without a decrease in economic activity (degrowth theory), as it is not proven that decarbonisation plans envisaged for emissions reductions, based on a combination of negative emissions and unprecedented technological change, can deliver a sufficient decoupling and that the rebound effect of the investments related to climate change may limit its benefits [[4], [5], [6]]. There is a middle approach that suggest the solution is taking into account the ecological limits and targeting GHG reduction, even if this may have a negative impact on GDP on short and medium term, to be reversed in the long run [7].

Even if historically the economic growth has been associated with an increase in GHG emissions, significant developments took place as new policies were implemented. From 1995 to 2018, global CO2 emissions per unit of GDP were declining, with a global decoupling rate of -1.8% annually (-3.4% in case of EU) [[1]]. Still, annual global GHG emissions kept rising (62% higher in 2019 compared to 1990 and 4% higher compared to 2015 level) until the COVID-19 crisis brought a significant drop. Global CO2 emissions declined by 5.8% in 2020 - the largest ever decline and almost five times greater than the 2009 decline that followed the global financial crisis [8]. In 2021 though, the GHG emissions rebounded to reach their highest ever annual level.

In the same vein, the energy-related emissions have decoupled from economic growth over the last two years in 21 countries(Austria, Bulgaria, Czeck Republic, Denmark, Finland, France, Germany, Hungary, Ireland, Netherlands, Portugal, Romania, Slovakia, Spain, Sweeden, Sweetherland, Ukraine, United Kingdom, United States) with an average change in the industry share in GDP of 3percentage points reduction over the period, and an average CO2 reduction of 15% [9]. As the economic literature indicate [3], the drivers of these developments are also very different: from reducing the industrial sector share of the economies (more than 90% of the 21 countries mentioned above that sow a change in their final energy consumption) to ambitious climate policies at national level (as in USA, Sweden or Denmark). Some authors [[10], [11]] have demonstrated the link between GHG and energy consumption, while others [12] indicated that energy intensity was the main determinant of GHG emission growth in the EU.

However, we cannot have a clear picture on decoupling, without taking into account that in developed countries, GHG emissions attributable to consumption may be higher than territorial emissions due to the fact that a lot of goods are produced elsewhere. The difference between territorial and consumption emissions has generally increased from 1990 to around 2005 and remained relatively stable afterwards [13]. At the same time, the development level is determinant for the impact of the GDP increase on GHG emissions: in countries that reached the Kuznets turning point, the GDP increase will positively impact the level of GHG [[14], [15]], and the direction and relationship between GDP and GHG differs from a country to another [16].

EU is a global leader in implementing ambitious climate change policies as well as investments in new technologies. Its first actions date back in 1990s, immediately after the first IPCC report was issued [17], when the EU leaders agreed to stabilise GHG emissions at 1990 levels by 2000. At the beginning, the actions were in the field of energy efficiency (Specific Actions for Vigorous Energy Efficiency agreed in 1991) and renewables (ATENER programme introduced, in 1993, an indicative target of 8% energy supply from renewables by 2005).

The Kyoto Protocol (agreed in December 1997) pushed the global community to take action, as the industrialized countries committed to a reduction of 8% reductions of GHG emissions during the 2008-2012. The EU responded by launching, in 2000, the European Climate Change Programme, with actions in GHG emissions reductions, energy efficiency and renewable energy.

The EU ambition increased in early 2000, once the 2020 energy and climate package was adopted [18], setting three key targets: 20% cut in greenhouse gas emissions (from 1990 levels), reaching 20% of EU energy consumption from renewables and 20% improvement in energy efficiency. Preliminary estimates indicate the full achievement - and even overachievement - of Europe's 20-20-20 goals for climate change mitigation, renewable energy deployment and energy efficiency gains [19].

Assessing the direct link between GDP and GHG emissions in this time lap is a difficult task, as several shocks impacted the economic activity: the 2008 global financial crisis, the 2019 pandemics. There are studies that demonstrated that financial crisis did have an impact on the effects produced on GHG by the economic growth, particularly in the countries where the economic growth rate was most volatile [14].

The period 2021-2030 was initially covered by the 2030 climate and energy framework [20], having as key targets: at least 40% cuts in greenhouse gas emissions (from 1990 levels), at least 32% share for renewable energy and at least 32.5% improvement in energy efficiency. Since 2019, the EU plan [21] targets no net emissions of greenhouse gases by 2050 and economic growth decoupled from resource use. Some estimations indicate that, in order to support the EU achievement of net-zero emissions goal, a decoupling rate of -9.4% is needed, so that the decoupling is still weak [[1], [12]]. Time will decide if the technological progress and the political ambition will be able to deliver on this goal.

Increase in renewables is one key policy option for European Union in fighting climate change. Indeed, studies prove that renewables help reducing GHG [22], even when the entire lifecycle is being assessed [23]. Indeed, the solution is not perfect. For example, some authors [24] show that solar might reduce aggregated risks, but it may increase climate risks for some regions, while others [25] indicate that renewable energy does not mean zero carbon, as emissions from manufacturing and installing technologies, as well as a mismatch between power consumption and renewable generation have an impact.

Circular economy and recycling play their role in climate change mitigation, but the studies indicate a mixed picture as regards the impact and demonstrate the complexity of this issue. At EU level, waste policy aims to protect the environment and human health and help the EU's transition to a circular economy [26] and it sets objectives and targets to improve waste management, stimulate innovation in recycling and limit landfilling and touches upon several policy areas: packaging, biodegradable, construction and demolition, batteries and accumulators, mining, sewage just to name few.

As previously mentioned, efforts were undertaken to quantify the GHG impacts of material recycling. While some authors [[15], [27]] demonstrated that recycling can play an important role in reducing the GHG impact of waste management, others [22] obtained mixed results in their study of the EU member states, thus showing that different development stage and policy options are determinant.

The literature in this field also propose other potential solutions to the challenges related to climate change / the growth of greenhouse gas emissions, as follows: (i) promoting alternative fuels in the maritime sector (such as biodiesels instead of petroleum fuels) [28]; (ii) including education on recycling as a compulsory primary school subject [[29], [30]]; (iii) reducing air and car travel, a global carbon tax and promoting the shift to the vegetarianism among people [31]; (iv) using stratospheric aerosol geoengineering to address climate change issues [32]; boosting building renovation [33]; supporting reforestation [34].

In sum, the main relevant factors for GHG developments that have been widely explored until now are related to: GDP per capita, the share of polluting industry, energy taxes, fuel mix and the use of biofuels, population, environmental pricing, final energy consumption, renewable energy, energy efficiency, expenditures on R&D, number of polluting vehicles, number of tourists and waste [[12], [14], [15], [35], [36]]. Taking into account the above mentioned factors, we will focus our paper on only a small part of these to avoid multicollinearity issues, the data being described in the material and methods section.

### III. MATERIALS AND METHODS

In this section, we have described the data and the methods used to assess the relationship between greenhouse gas emissions and its drivers over the period 2010-2019 in a Panel comprising data for the EU-27 Member States (Eurostat data - [37]). We used this period to limit the effect of the crisis shocks on the estimation. Even if this period corresponds to a EU-28 format, we dropped UK from our analysis, since in its case, we did not find 2019 data for some indicators.

Further, we have decided the model specification and method depending on the stationarity tests results. In this context, we have used Schwarz information criterion for setting the appropriate lag and five tests to decide if our data are stationary at level, at first difference or at second difference, as follows:
- Levin, Lin and Chu t*;
- Breitung t-stat;
- Im, Pesaran and Shin W-stat;
- ADF-Fisher Chi-square;
- PP-Fisher Chi-square.

After assessing the stationarity tests results, we concluded that our data are stationary at level and at first difference in some cases, which requires adding lags for the dynamic variables (which are stationary after computing the first difference) according to the economic theory but also to the significant coefficients. In the process of identifying the number of lags, we used five criteria (sequential modified LR test statistic; final prediction error; Akaike information criterion; Schwarz information criterion and Hannan-Quinn information criterion) and we found that 2 lags are appropriate to our model. However, our estimation indicated that some impact coefficients are not significant and we removed those from our analysis.

In addition, we expressed some data in natural logarithm form in order to better catch the growth dynamics of the indicators and to tackle the heteroscedasticity issues that may appear within the model. Following the transformations indicated above, we applied the Redundant Fixed Effects Likelihood Ratio (prob. - 0.00) test to check the compatibility with the Fixed Effects Model in order to set fixed effects for every year taken into consideration. This approach also allows an homogenous treatment of every year to limit the specific effects of yearly developments on the final results.

Therefore, we have applied the Panel EGLS method with period fixed effects (weighted by Period SUR option to ex-ante address the issues related to heteroscedasticity and cross-sectional dependence) on equation (1).

$$logghgpercap_{it} = \alpha_0 + \alpha_1 logrgdppercap\_t - 1_{it} + \alpha_2 renewable_{it} + \alpha_3 loghhenergyconspercap_{it} + \alpha_4 logwastepercap\_t - 2_{it} + \varepsilon_t \quad (1)$$

, where:

$logghgpercap$ is the natural logarithm of the greenhouse gas emissions per capita (expressed in CO2 equivalent), $logrgdppercap\_t-1$ reflects the natural logarithm of the real GDP per capita lagged by one year, $renewable$ represents the share of renewable energy in gross final energy consumption, $loghhenergyconspercap$ reflects the natural logarithm of final energy consumption in households per capita (expressed in kilogram of oil equivalent - KGOE), $logwastepercap\_t-2$ refers to the natural logarithm of the waste generated (expressed in kilograms) per capita lagged by 2 years, $\varepsilon_t$ represents the residuals, $\alpha_{0-4}$ are the estimated parameters, while $i$ reflects the number of countries and $t$ is the number of periods.

Further, we have added 9 dummy variables to the equation (1) to ensure a proper estimation of the fixed effects model (number of periods - 1), as we have indicated in equation (2):

$$logghgpercap_{it} = \beta_0 + \beta_1 logrgdppercap\_t-1_{it} + \beta_2 renewable_{it} + \beta_3 loghhenergyconspercap_{it} + \beta_4 logwastepercap\_t-2_{it} + \rho_1 dummy_1 + \ldots + \rho_9 dummy_9 + \mu_t \quad (2)$$

, where:

$\rho_1, \ldots, \rho_9$ represents the intercepts of the dummies, $\beta_0, \ldots, \beta_4$ are the new impact coefficients and $\mu_t$ is the adjusted residuals series.

Finally, we analysed the feasibility of the model by checking the following hypotheses: (i) statistical validity of the model - Fisher test; (ii) significance of the coefficients; (iii) normal distribution of the residuals - Jarque-Bera test and their null mean; (iv) cross-section dependence - Breusch-Pagan LM, Pesaran scaled LM and Pesaran tests; (v) multicollinearity - Variance Inlation Factors test; (vi) linearity of the model - R-squared.

## IV. RESULTS

In this section, we described the main results obtained according to the methodology used. At EU level, the highest level of greenhouse gases per capita in 2019 were found in LU (20.3 tonnes of CO2 equivalent per capita), IE (12.8 tonnes of CO2 equivalent per capita) and CZ (11.7 tonnes of CO2 equivalent per capita), while the lowest level were registered in SE (5.2 tonnes of CO2 equivalent per capita), MT (5.3 tonnes of CO2 equivalent per capita) and RO (5.9 tonnes of CO2 equivalent per capita). There is also a need to mention the countries that register the most significant progress in terms of GHG emissions reduction in the period of 2010-2019 are LU (-6.3 tonnes of CO2 equivalent per capita), EE (-4.8 tonnes of CO2 equivalent per capita) and FI (-4.3 tonnes of CO2 equivalent per capita).

In the analysed period (2010-2019), greenhouse gases per capita decreased by 1.2 tonnes of CO2 equivalent per capita continuing the regressive path since 1990, this being also a result of the EU policy in the environment field. However, addressing greenhouse gas emissions issues remains one of the main priorities in climate field to tackle global warming challenges and may be the unique solution to avoid the sixth extinction on long-run. In this context, studying the effects of the GHG emissions drivers has a particular importance for the process of identifying new solutions to tackle climate change. Therefore, we started by interpreting the estimated effects according to the methodology already presented above.

According to Table 1, an increase by 1% in real GDP per capita lagged by one year led to a hike in the greenhouse gas emissions per capita by 0.187%. This is mainly caused by the fact that the global economies are still based on polluting activities, including on the industrial sector even if, at EU-27 level, the share in GDP of the gross value added specific to the industrial sector marginally decreased in 2020, compared to 2000. However, this does not mean if we will make all efforts to reduce the industrial sector dimension, we will achieve the planned climate objectives. In addition, the industrial sector has also a particular importance for the economic resilience, since the businesses operating in the tertiary sector change more rapidly its location in the case of an economic shock, compared to the businesses operating in the industrial sector.

**Table 1.** Results of the model.

| Variables (dependent variable - logghgpercap) | Model Coefficients / Std. Error |
|---|---|
| logrgdppercap(-1) | 0.186541* (0.036509) |
| renewable | -0.014396* (0.001485) |
| lohhhenergyconspercap | 0.328649* (0.028504) |
| logwastepercap(-2) | 0.057937** (0.029292) |
| Constant | -1.795949* (0.381643) |

*Greenhouse Gas Emissions and its Main Drivers: a Panel Assessment for EU-27 Member States*

| Model tests | |
|---|---|
| R-squared | 0.726682 |
| F-statistic (prob.) | 0.0000 (p<.05) |
| Observations | 270 (10 / cross-section) |
| Cross-section dependence test | |
| Breusch-Pagan LM (prob.) | 0.8969 (p>.05) |
| Pesaran scaled LM (prob.) | 0.0234 (p<.05) |
| Pesaran CD (prob.) | 0.0716 (p>.05) |
| Normality of the residuals test | |
| Jarque-Bera (prob.) | 0.1016 (p>.05) |
| Standardized residuals - mean | -0.0000000000000019 |
| Multicollinearity test - Variance Inflation Factors - Centered VIF coefficients | |
| logrgdppercap(-1) | 1.145421 |
| renewable | 1.080022 |
| lohhhenergyconspercap | 1.185811 |
| logwastepercap(-2) | 1.007024 |

**Note:** *significant at 1%, **significant at 5%; standard errors in parentheses.
*Source: Own calculations using Eviews 10.0.*

With a view to the share of renewable energy in gross final energy consumption, we found that an increase in this indicator by 1 percentage point has led to a reduction in the greenhouse gas emissions per capita by 0.014%. This also provide additional evidence supporting the hypothesis that the renewable energy is important, but should be accompanied by other environmentally friendly reforms to support a significant reduction of the greenhouse gas emissions, taking into account its low impact on GHG reduction. In addition, it worth to be mentioned that 26 Member States (excepting France which registered a share of the renewable energy in gross final energy consumption equal to 19.10%, a lower level to the national target of 23%) achieved their national targets resulted from the Europe 2020 Strategy in the field of renewable energy as it can be observed in Fig. 1. However, when assessing the greenhouse gas emissions reported for 2018 (Fig. 2), we found that only 12 Member States succeed to reduce their emissions by at least 20% compared to 1990 level. Even if 2019 and 2020 data were not available, the outlooks for achieving the GHG emissions objective is less favourable than the one related to the renewable energy share in 2018.

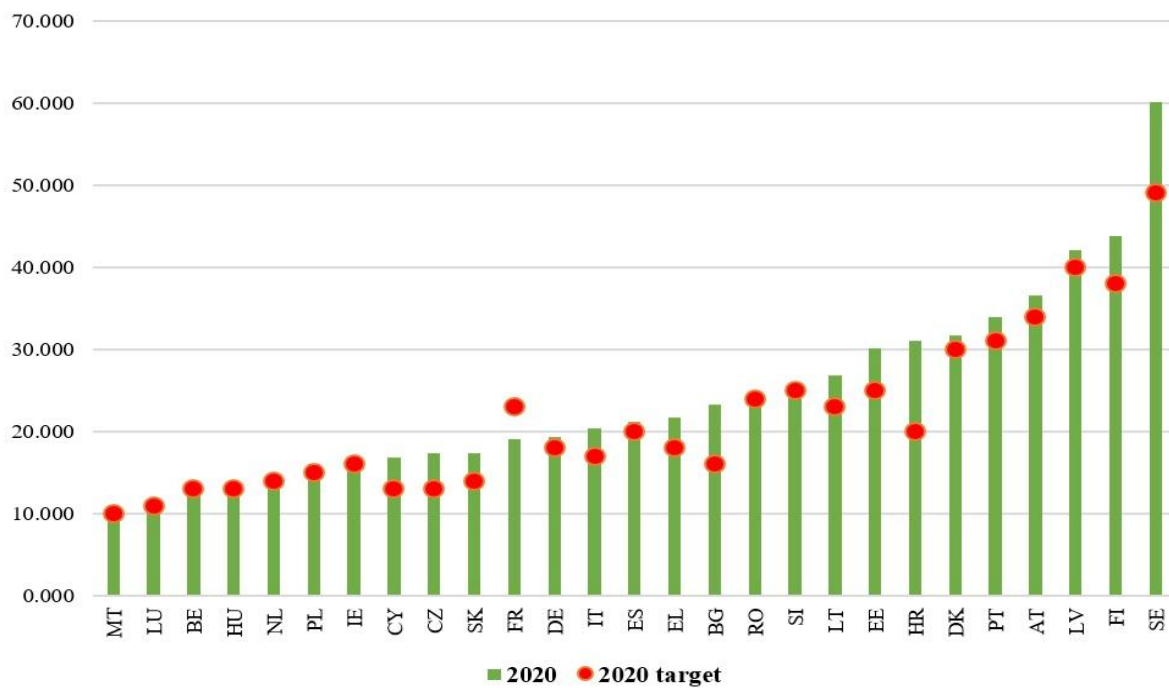

**Figure 1.** share of renewable energy in gross final energy consumption in EU-27 (2020 level vs. 2020 target).
*Source: Own performings using Microsoft Office Excel 2016.*

*Greenhouse Gas Emissions and its Main Drivers: a Panel Assessment for EU-27 Member States*

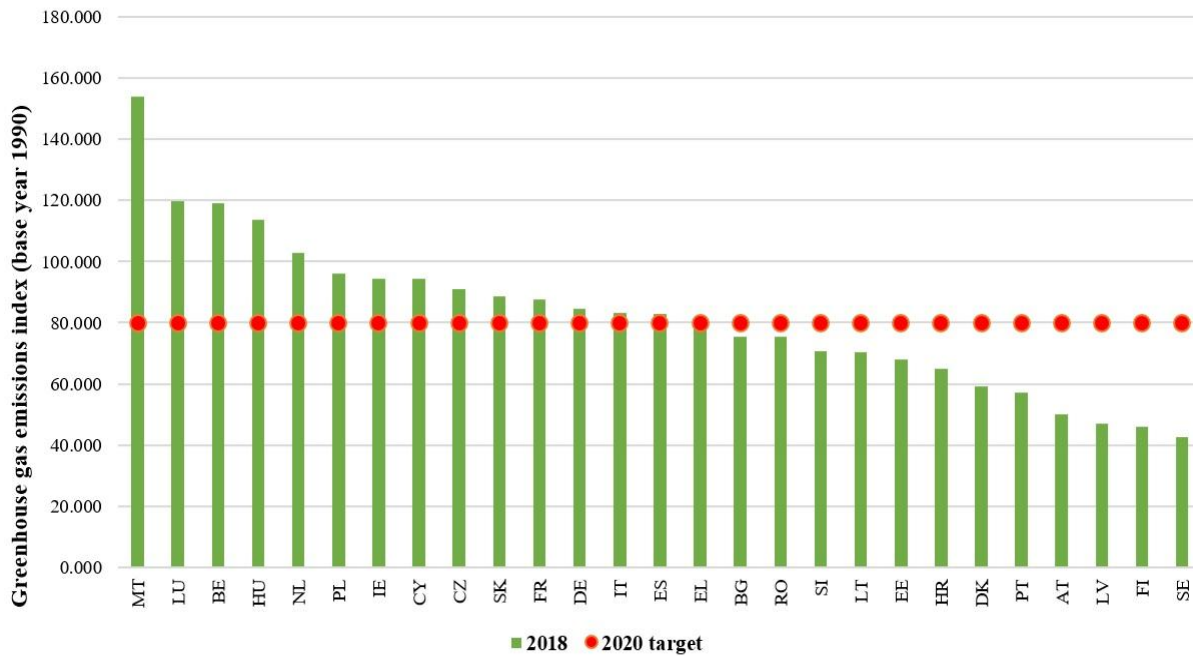

**Figure 2.** greenhouse gas emissions reduction in EU-27 compared to 1990 level (2018 level vs. 2020 target)
*Source: Own performings using Microsoft Office Excel 2016*

Further, we found that an increase of the final energy consumption in households per capita (expressed in kilogram of oil equivalent - KGOE) by 1% caused a growth of the greenhouse gas emissions per capita by 0.329%, this also being the highest impact among the effects of the GHG emissions drivers we assessed. As it can be observed in Fig. 3, at EU-27 level, energy consumption in households are generally composed by environmentally harmful fuels, only 20% of it being related to renewable sources or biofuels. This argue the positive relationship between households energy consumption and greenhouse gas emissions.

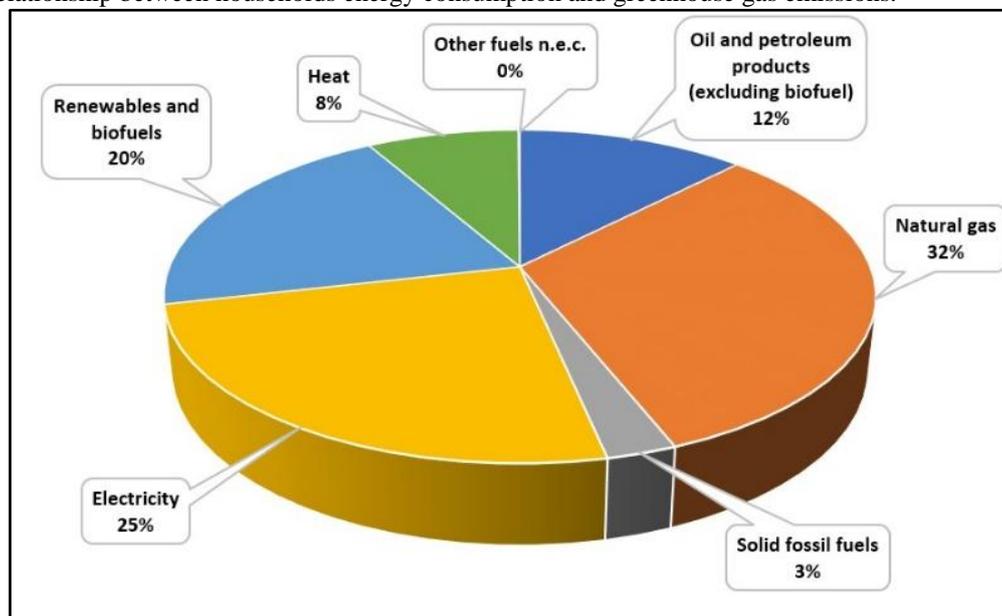

**Figure 3.** the structure of households final energy consumption in EU (2020 data)
*Source: Own performings using Microsoft Office Excel 2016.*

Finally, we found that an increase by 1% in waste generated per capita has led to a hike in the greenhouse gas emissions per capita by 0.057% after two years when the shock occurs. Waste is a significant challenge that world is facing with since waste decomposition takes several years to, but it increases pollution from the start of this process. In this respect, promoting recycling behaviour remains a good way forward to tackle excessive waste generation, but also high levels of GHG emissions.

Next, we presented the results of the main tests performed to check the feasibility of the model. According to Table 1, all estimators are statistically significant at 5%, three drivers being significant at 1% (real

GDP per capita lagged by one year, the share of renewable energy in final energy consumption and final households energy consumption). The results indicate a R-squared value of 72.66% that confirms the appropriate selection of the regressors since the evolution of the greenhouse gas emissions is explained in a proportion of 72% by its drivers. In addition, the probability of the Fisher test (0.00%) confirms the statistical validity of the model, while Breusch-Pagan LM (prob. 89.69%) and Pesaran CD (prob. 7.16%) tests validate the hypothesis of no cross-section dependence. To verify the normality of the residuals, we used the Jarque-Bera test, which confirmed the residuals are normally distributed (prob. 10.16% > 5%). Finally, we checked the presence of the multicollinearity in our model using the Variance Inflation Factors test (VIF), which confirmed its absence taking into account that the centered VIF coefficients are lower than 4 this being an appropriate threshold for identifying even a low kind of multicollinearity.

## V.  CONCLUSION AND DISCUSSIONS

Our paper assesses the effects of greenhouse gas emissions drivers in EU-27 over the period 2010-2019. We found a positive relationship between three independent variables (real GDP per capita, households final consumption per capita and waste generation per capita) and greenhouse gas emissions per capita, while the effect of the share of renewable energy in gross final energy consumption on the dependent variable proved to be negative, but quite low (the increase in this indicator by 1 percentage point has led to a reduction in the greenhouse gas emissions per capita by 0.014% over the analysed period). Renewable energy is an important solution to tackle the increasing path of greenhouse gas emissions, but its lower share in the households energy consumption (20%) reduce its capacity to reduce emissions.

In addition, we demonstrate that the main challenge that affects greenhouse gas emissions is related to the structure of households energy consumption, which is generally composed by environmentally harmful fuels (an increase of the households energy consumption per capita by 1% caused a growth of the greenhouse gas emissions per capita by 0.329%). Moreover, we found that an increase by 1% of the real GDP per capita has led to a hike in the greenhouse gas emissions per capita by 0.187%. Our findings call the need to make greater efforts to support the shift to a green economy based on a higher energy efficiency, but also on the less polluting activities. In this respect, technology may play a significant role.

Further, we concluded that an increase by 1% of waste generation per capita is responsible for a growth of greenhouse gas emissions by 0.057%. In this context, we propose increasing the quality of waste management and promoting the recycling behaviour among the individuals. This remains an important good way forward, since daily per capita waste generation is projected to grow by 19% in high-income countries and with at least 40% in low and middle-income countries, until 2050 [38]. Our finding is somehow in line with World Bank data indicating that 5% of the global greenhouse gas emissions, but the conclusions associated to our research are limited to EU-27 Member States, while the results should be interpreted in marginal terms.

Our paper also identifies some key policy recommendations to tackle greenhouse gas emissions / climate change, in line with the scientific literature in this field and to the European Union actions in climate field.

As the households energy consumption proved to be the main challenge, we suggest (i) establishing a carbon tax, while maintaining some kind of flexibility for inclusiveness purposes; (ii) promoting the use of environment friendly vehicles such as bicycles, electric scooters, electric / hybrid cars, electric common transport vehicles; (iii) applying cash benefits for consumers buying energy efficient products; (iv) implementing buildings energy renovation reforms in a higher extent.

Although the negative impact of renewables on GHG is quite small according to the results of our model, we consider that increasing the share of renewable energy in final energy consumption need to be part of the solution. In addition, as waste generation per capita do impact the GHG emissions, promoting recycling behaviour and improving the quality of waste management, as well as improving the quality of water management should be considered by the policy makers.

At the same time, other solutions proposed by the literature should be taken into account: (v) promoting ecological agriculture; (vi) integrating green budgeting principles in governmental budgetary processes; (vii) supporting reforestation; (viii) introducing learning disciplines focused on climate change issues in all levels of education; (xi) making greater efforts to exploring other potential solutions to tackle climate change which should be implemented in parallel with the above mentioned recommendations (e.g.: stratospheric Aerosol Injection).

The limitations of our papers are related to the fact that we used a general Panel approach at EU-27 level, not a country-specific approach. In this context, the results obtained are relevant only for the EU as a whole and should not been adapted to the country level. In addition, even if the analysed period (2010-2019) corresponds to the EU-28 format, we dropped UK from our analysis, since in its case, we did not find 2019 data for some indicators.

Further research directions include extending the analysis by adding other determinants of greenhouse gas emissions in the assessment, without triggering multicolliniarity issues, and also by addressing particular effects at country-group level, taking into account country-specificities. In addition, we will extend our research by analysing the bidirectional relationship between environmental degradation and economic development.

## REFERENCES

[1] K. Lenaerts, S. Tagliapietra, and G.B. Wolff, Can climate change be tackled without ditching economic growth?, *Bruegel Working paper*, no. 10/2021.
[2] M. Tucker, Carbon dioxide emissions and global GDP, *Ecological Economics, Vol. 15(3), 1995,* 215-223.
[3] A. Nate, The roads to decoupling: 21 countries are reducing carbon emissions while growing GDP, *World Resources Institute Business Center communication*, 2016.
[4] L. Keyßer, and M. Lenzen, 1.5°C degrowth scenario suggest the need for new mitigation pathways, *Nature Communications*, Vol. 12(1), 2021, Article 2676.
[5] J. Eastin, R. Grundman, and A. Prakash, The two limits debates: limits to growth and climate change, *Futures*, Vol. 43(1), 2011, 16-26.
[6] M. Antal, and J.C.J.M. van den Bergh, Green growth and climate change: conceptual and empirical considerations, *Climate Policy*, Vol. 16(2), 2016, 165-177.
[7] J.J. Fitzpatrick, Target ecological limits and not economic growth, *World*, Vol. 1(2), 2020, 135-148.
[8] International Energy Agency, Global Energy Review 2021 - Assessing the effects of economic recoveries on global energy demand and $CO_2$ emissions in 2021, *IEA report*.
[9] International Energy Agency, Decoupling of global emissions and economic growth confirmed, *IEA communication*.
[10] M.A. Khan, M.Z. Khan, K. Zaman, and L. Naz, Global estimates of energy consumption and greenhouse gas emissions, *Renewable and Sustainable Energy Reviews,* Vol. 29(C), 2014, 336-344.
[11] S.A. Sarkodie, and V. Strezov, Effect of foreign direct investments, economic development and energy consumption on greenhouse gas emissions in developing countries, *Science of the Total Environment*, Vol. 646, 2019, 862-871.
[12] M. Gonzales-Sanchez, and J. L. Martin-Ortega, Greenhouse gas emissions growth in Europe: a comparative analysis of determinants, *Sustainability*, Vol. 12(3), 2020, Article 1012.
[13] P. Friedlingstein, M. O'Sullivan, M.W. Jones, R.M. Andrew, J. Hauck, A. Olsen, G.P. Peters, W. Peters, J. Pongratz, S. Sitch, C. Le Quéré, J.G. Canadell, P. Ciais, R.B. Jackson, S. Alin, L.E.O.C. Aragão, A. Arneth, V. Arora, N.R. Bates, M. Becker, A. Benoit-Cattin, H.C. Bittig, L. Bopp, S. Bultan, N. Chandra, F. Chevallier, L.P. Chini, W. Evans, L. Florentie, P.M. Forster, T. Gasser, M. Gehlen, D. Gilfillan, T. Gkritzalis, L. Gregor, N. Gruber, I. Harris, K. Hartung, V. Haverd, R.A. Houghton, T. Ilyina, A.K. Jain, E. Joetzjer, K. Kadono, E. Kato, V. Kitidis, J.I. Korsbakken, P. Landschützer, N. Lefèvre, A. Lenton, S. Lienert, Z. Liu, D. Lombardozzi, G. Marland, N. Metzl, D.R. Munro, J.E.M.S. Nabel, S. I. Nakaoka, Y. Niwa, K. O'Brien, T. Ono, P.I. Palmer, D. Pierrot, B. Poulter, L. Resplandy, E. Robertson, C. Rödenbeck, J. Schwinger, R. Séférian, I. Skjelvan, A.J.P. Smith, A.J. Sutton, T. Tanhua, P.P. Tans, H. Tian, B. Tilbrook, G. van der Werf, N. Vuichard, A.P. Walker, R. Wanninkhof, A.J. Watson, D. Willis, A.J. Wiltshire, W. Yuan, X. Yue, and S. Zaehle, Global carbon budget 2020, *Earth System Science Data,* Vol. 12, 2020, pp. 3269-3340.
[14] G. Lapinskiene, K. Peleckis, and M. Radavicius, Economic development and greenhouse gas emissions in the European Union countries. *Journal of Business Economics & Management,* Vol. 16(6), 2015, 1109-1123.
[15] T. Vasylieva, O. Lyulyov, Y. Bilan, and D. Streimikiene, Sustainable economic development and greenhouse gas emissions: the dynamic impact of renewable energy consumption, GDP, and corruption, *Energies*, Vol. 12(17), 2019, Article 3289.
[16] M. Păunică, A. Manole, C. Motofei, and G.L. Tănase, Causality analysis between economic development and greenhouse gases emissions - an European Union perspective, *Economic Computation and Economic Cybernetics Studies and Research*, Vol. 54(4), 2020, 187-202.
[17] Climate Policy Info Hub, European Climate Policy - History and State of Play, *Climate Policy Info Hub communication*.
[18] European Commission, Climate Action: 2020 climate & energy package, *European Commission website information.*
[19] European Environment Agency, Trends and Projections in Europe 2021, *EEA Report No 13/2021*.
[20] European Commission, Climate Action: 2030 climate & energy framework, *European Commission website information.*
[21] European Commission. Communication From The Commission: The European Green Deal, *COM(2019) 640 final*, Brussels.

[22] Y. Bayar, M.D. Gavriletea, Ș. Sauer, and D. Păun, Impact of municipal waste recycling and renewable energy consumption on $CO_2$ emissions across the European Union (EU) member countries, *Sustainability*, Vol. 13(2), 2021, Article 656.

[23] Y.N. Amponsah, M. Troldborg, B. Kington, I. Aalders, and R.L. Hough, Greenhouse gas emissions from renewable energy sources: a review of lifecycle considerations, *Renewable and Sustainable Energy Reviews,* Vol. 39(C), 2014, 461-475.

[24] P. Irvine, K. Emanuel, J. He, L.W. Horowitz, G. Vecchi, and D. Keith, Halving warming with idealized solar geoengineering moderates key climate hazards, *Nature Climate Change*, Vol. 9, 2019, 295-299.

[25] V. Xia, When 100% renewable energy doesn't mean zero carbon, *Stanford Earth Matters magazine*, 2019.

[26] European Commission, Waste and Recycling, *European Commission website information.*

[27] D.A. Turner, I.D.Williams, S. Kemp, Greenhouse gas emission factors for recycling of source-segregated waste materials, *Resources, Conservation and Recycling,* Vol. 105(A), 2015, 186-197.

[28] Y. Wang, and L.A. Wright, A comparative review of alternative fuels for the maritime sector: economic, technology and policy challenges for clean energy implementation, *World*, Vol. 2(4), 2021, 458-481.

[29] E. Altikolatsi, E. Karasmanaki, A. Parissi, and G. Tsantopoulos, Exploring the factors affecting the recycling behavior of primary school students, *World*, Vol. 2(3), 2021, 334-350.

[30] E. Eliam, Climate change education: the problem with walking away from disciplines, *Studies in Science Education*, 2022, 1-34.

[31] P. Moriaty, and D. Honnery, New approaches for ecological and social sustainability in a post-pandemic world, *World*, Vol. 1(3), 2020, 191-204.

[32] P.J. Irvine, and D.W. Keith, Halving warming with stratospheric aerosol geongineering moderates policy-relevant climate hazards, *Environmental Research Letters*, Vol. 15(4), 2020, Article 044011.

[33] European Commission, Communication from the Commission to the European Parliament, The Council, The European Economic and Social Committee and the Committee of the Regions: A Renovation Wave for Europe - Greening Our Buildings, Creating Jobs, Improving Lives, *COM(2020) 662 final*, Brussels.

[34] B. Waring, M. Neumann, I.C. Prentice, M. Adams, P. Smith, and M. Siegert, Forests and decarbonization - roles of natural and planted Forests, *Frontiers in Forests and GlobalChange*, Vol. 3(58), 2020.

[35] M. Ziolo, K. Kluza, and A. Spoz, Impact of sustainable financial and economic development on greehnouse gas emission in the developed and converging economies, *Energies*, Vol. 12(3), 2019, Article 4514.

[36] A. Khan, S. Bibi, L. Ardito, J. Lyu, H. Hayat, and A.M. Arif, Revisiting the dynamics of tourism, economic growth, and environmental polluants in the emerging economies - sustainable tourism policy implications, *Sustainability*, Vol. 12(6), 2020, Article 2533.

[37] Eurostat, *Eurostat database*, 2022.

[38] World Bank, Trends in Solid Waste Management, *World Bank website information*.

*Corresponding Author: Ionuț Jianu[1]
[1](Department of Economics and Economic Policies, Bucharest University of Economic Studies, Romania)*